# Ultrahigh thermoelectric performance of Janus α-STe$_2$ and α-SeTe$_2$ monolayers

Gang Liu[1*], Aiqing Guo[1], Fengli Cao[1], Weiwei Ju[1], Zhaowu Wang[1], Hui Wang[1], Guo-Ling Li[2], Zhibin Gao[3*]

[1]School of Physics and Engineering, Henan University of Science and Technology, Luoyang 471023, People's Republic of China

[2]Chemistry and Chemical Engineering Guangdong Laboratory, Shantou 515063, People's Republic of China

[3]State Key Laboratory for Mechanical Behavior of Materials, Xi'an Jiaotong University, Xi'an 710049, People's Republic of China.

## Abstract

Combined with first-principles calculations and semiclassical Boltzmann transport theory, Janus α-STe$_2$ and α-SeTe$_2$ monolayers are investigated systematically. Janus α-STe$_2$ and α-SeTe$_2$ monolayers are indirect semiconductors with band gaps of 1.20 and 0.96 eV. It is found they possess ultrahigh figure of merit (*ZT*) values of 3.9 and 4.4 at 500 K, much higher than that of the pristine α-Te monolayer (2.8). The higher ZT originates from Janus structures reduce lattice thermal conductivities remarkably compared with pristine α-Te monolayer. The much higher phonon anharmonicity in Janus monolayers leads to significant lower lattice thermal conductivity. It is also found electronic thermal conductivity can play an important role in thermoelectric efficiency for the materials with quite low lattice thermal conductivity. This work suggests the potential applications of Janus α-STe$_2$ and α-SeTe$_2$ monolayers as thermoelectric materials and highlights Janus structure as an effective way to enhance thermoelectric performance.

---
*Corresponding author: Gang Liu, Email: liugang8105@haust.edu.cn
*Corresponding author: Zhibin Gao, Email: zhibin.gao@xjtu.edu.cn

## Introduction

During the past several decades, thermoelectric (TE) materials attract much attention as they can convert waste heat into useful electrical power directly. Generally, a dimensionless figure of merit *ZT* can be used to evaluate the conversion efficiency, which can be expressed as $ZT = S^2\sigma T/(\kappa_e + \kappa_L)$, where the *S*, *σ*, *T*, $\kappa_e$ and $\kappa_L$ are Seebeck coefficient, electrical conductivity, absolute temperature, electronic thermal conductivity, and lattice thermal conductivity, respectively. Moreover, $S^2\sigma$ is the power factor (PF). A high *ZT* value requires a high $S^2\sigma$ and low thermal conductivity ($\kappa_e + \kappa_L$). Unfortunately, the parameters which determine the *ZT* value are usually interrelated. For instance, *S* and *σ* generally behave in an opposite trend, greatly complicating the optimizing of *ZT* values [1-3]. Usually, a TE material should possess a narrow band gap, due to excellent electronic transport properties [4, 5]. However, it is reported recently that materials with wide band gaps can also have excellent TE performance due to high motilities and low lattice thermal conductivities [6-8]. Various theoretical and experimental investigations suggest reducing the dimensionality can improve the *ZT* value of the TE materials, as the quantum confinement effect and the interface/surface scattering effect [9-12].

Since the successful exfoliation of graphene in 2004 [13, 14], a new field in the study of two-dimensional (2D) materials has been opened up. Numerous 2D materials have been investigated theoretically and experimentally, as their unique physical and chemical properties [15-22]. Recently, motivated by the successful synthesis of tellurene (2D tellurium) [22], the 2D materials of group VI attract much attention, and various 2D materials of Te and Se have been synthesized experimentally [22-28]. It is found the tellurene possesses high carrier mobility on the order of $10^3$ cm$^2$ V$^{-1}$s$^{-1}$ [22], outstanding air stability [29], and a high on/off ratio on the order of $10^6$ [24], leading to potential applications in field-effect transistor (FET), photodetectors, and sensors. Extraordinary electronic transport properties have also been found for few-layer tellurene [24, 30, 31]. Furthermore, it also shows good potential applications for thermoelectric devices with high thermoelectric performance [11, 32, 33].

Recently, Janus 2D materials become a hot issue due to the mirror asymmetry in the structure and resultant distinct properties. The Janus transition metal dichalcogenides (TMDs) have been investigated theoretically and experimentally [34-37]. Investigations suggest that the Janus monolayer has promising applications in various fields such as electronics, optoelectronics, photocatalysts, and gas sensing [38-41]. In addition, theoretical researches find Janus monolayers usually reduce $\kappa_L$ obviously compared with the pristine monolayers, such as Janus MoSSe, ZrSSe,

PtSSe, and SnSSe [42-45]. Thus, the Janus structure may possess higher ZT than the pristine material due to its low thermal conductivity [46].

In this work, we have investigated the TE properties of Janus α-STe$_2$ and α-SeTe$_2$ monolayers based on first principles. For comparison, the pristine α-Te monolayer was investigated as well. The dynamic and thermal stability is confirmed by phonon dispersion and the *ab initio* molecular dynamics (AIMD) simulations. It is found Janus α-STe$_2$ and α-SeTe$_2$ monolayers possess ultra-high *ZT* values which are up to about 4.0 at 500 K, much higher than the maximum *ZT* value of about 2.8 for α-Te. The ultra-high *ZT* originates from the good electronic transport properties and ultra-low thermal conductivity. Janus structures reduce the electronic transport properties as the breaking of inversion symmetry, but also impair $\kappa_L$ significantly, leading to the enhancement of *ZT*. Furthermore, the importance of electronic thermal conductivity to TE efficiency is highlighted. The physical mechanisms of low $\kappa_L$ are also investigated. It is found the much higher phonon anharmonicity results in the lower $\kappa_L$ in Janus monolayers. And the broken inversion symmetry leads to the higher phonon anharmonicity in Janus monolayers hence lower $\kappa_L$.

## Computational Methods

Based on the density functional theory (DFT), all the first-principles calculations were performed by the Vienna *ab initio* simulation package (VASP) [47, 48]. The Perdew−Burke−Ernzerhof (PBE) generalized gradient approximation (GGA) was used as the exchange-correlation functional [49].

For structure optimization, a plane-wave kinetic energy cutoff of 600 eV was chosen, which is much higher than the maximum recommended cutoff to ensure the good convergence of the calculations. During the optimization, a total energy convergence criterion of $10^{-8}$ eV was chosen, while the force convergence criterion was $10^{-4}$ eV/Å. The Monkhorst–Pack [50] k-mesh of 15 × 15 × 1 was used to sample the Brillouin zone (BZ). The van der Waals (vdW) correction proposed by Grimme [51] was taken into consideration during the calculations. The effect of spin-orbit coupling (SOC) was also included due to the presence of heavy elements. As band gap is always underestimated with PBE, HSE06 [52] was employed to obtain accurate electronic band structure and density of states (DOS). To verify the thermal stability, the *ab initio* molecular dynamics (AIMD) simulations controlled by the Nose-Hoover thermostat [53, 54] have been performed for 5000 fs with a time step of 1 fs at 500 K. The electronic transport properties are implemented in the BoltzTraP2 code [55], where Boltzmann transport theory and relaxation time approximation (RTA) are utilized. The phonon dispersions were obtained by Phonopy [56]. AlmaBTE code was used to calculate the lattice thermal conductivity by solving the Boltzmann transport equation [57].

## Results and Discussions

The structure of the α-Te monolayer belongs to $\bar{P}3M1$ (164) symmetry group, similar to 1T-MoS$_2$. The Janus monolayers are built from the pristine α-Te by replacing Te atoms of the outer layer with S/Se atoms.

The optimized structures of three monolayers are shown in Fig. 1, including the α-Te. From the side views, it can be found the symmetry is broken as two surfaces of the Janus monolayer consist of different chalcogen atoms. The lattice constant $a$ and buckling height $d$ of optimized structures are listed in Table 1. The cohesive energies $E_c$ per atom for the three materials are also calculated, as listed in Table 1 as well. Cohesive energies $E_c$ is defined as: $E_c = \frac{E_{tot}-(mE_{Se}+nE_{Te})}{m+n}$, where $E_{tot}$ is the total energy of the material studied, while $E_{Se}$ and $E_{Te}$ are the ones of isolated Se and Te atoms. Here $m$ and $n$ are the numbers of Se and Te atoms in a primitive cell. The calculated $E_c$ are -2.38, -2.24, and -2.08 eV for α-STe$_2$, α-SeTe$_2$ and α-Te monolayers, respectively. It indicates the energetic stability of the Janus monolayers. Moreover, the AIMD simulations also confirm the thermal stability of the three monolayers up to 500 K, as shown in Fig. S1 of the Supporting Information.

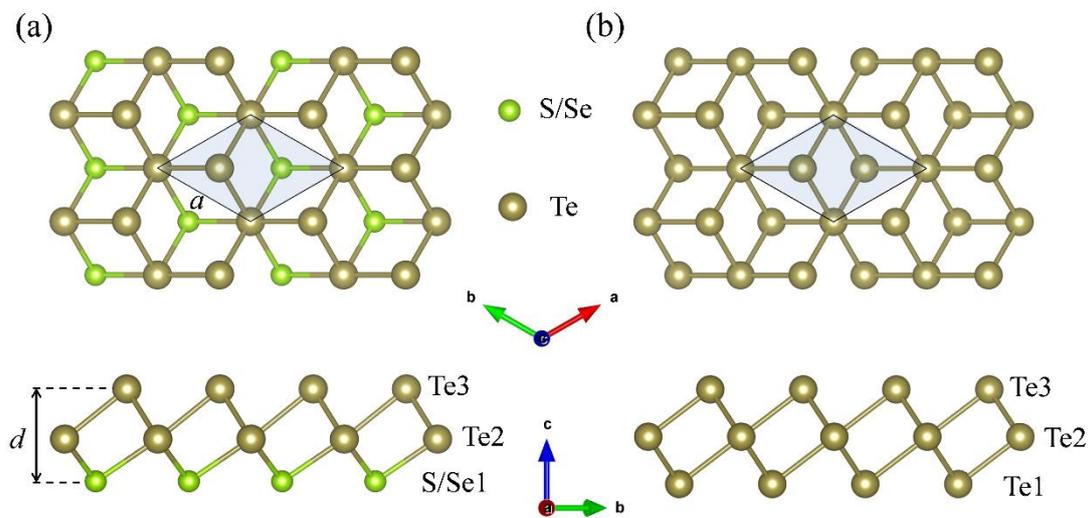

Fig. 1. Side and top views of Janus (a) α-S/SeTe$_2$ and (b) α-Te monolayers. The primitive cells are displayed by blue shading. The symbols S/Se/Te1, Te2, Te3 indicate atoms locate different layers.

Table 1. Lattice constant *a* and buckling height *d* of optimized structures, as well as the cohesive energies $E_c$ per an atom.

|  | α-STe$_2$ | α-SeTe$_2$ | α-Te |
|---|---|---|---|
| *a* (Å) | 4.03 | 4.05 | 4.15 |
| *d* (Å) | 3.34 | 3.49 | 3.67 |
| $E_c$ (eV) | -2.38 | -2.24 | -2.08 |

Electronic band structure plays an important role in TE performance. It should be noted that HSE06 hybrid functional including SOC effect (HSE+SOC) is necessary for accurate band structure of the materials containing heavy elements. For instance, the experimental value of the band gap is about 0.33 eV for bulk Te [58], while the calculated value is 0.31 eV with the HSE+SOC method [31], very close to the experimental one. For comparison, the PBE, PBE+SOC, and HSE without SOC methods lead to values different from the experimental value remarkably [31]. The band gaps with HSE+SOC, HSE, PBE+SOC, and PBE are listed in Table 2. For α-STe$_2$, α-SeTe$_2$, and α-Te monolayers, the calculated band gaps of HSE+SOC are indirect with values of about 1.22, 0.96, and 0.70 eV, respectively. The data are in good agreement with previous works [22, 33, 59]. Furthermore, the band gaps corresponding to other methods are significantly different from the ones of HSE+SOC. For instance, the gaps of HSE are 1.51, 1.21, and 1.12 eV, significantly larger than the ones of HSE+SOC. It also verifies that the SOC effect can't be neglected for the materials containing heavy Te element. The calculated electronic band structures and partial DOS (PDOS) of three monolayers with the method of HSE+SOC are shown in Fig. 2, while the band structures of HSE without SOC are exhibited as well. The conduction band minimum (CBM) is located at the Γ point for all the monolayers. However, the valence band maximum (VBM) is located in the Γ−M line for α-Te, while they are located in the Γ−K line for α-STe$_2$ and α-SeTe$_2$. On the whole, the shapes of band structures are similar, implying they may have similar values of effective masses overall for three monolayers, as listed in Table 3. However, the VBM of α-Te is located far from the ones of Janus monolayers, resulting in a much difference of $m^*$ of hole. Moreover, the valence bands near the Fermi level have relatively flat bands, showing the characteristic of high PF value, similar to Na$_x$CoO$_2$ and FeAs$_2$ [60-62]. In Fig. 2, it is also found SOC remarkably affects the shapes of band structure and the locations of VBM and CBM. The shapes of curves round VBM and CBM have great effect on $m^*$. Thus, we choose the data of HSE+SOC to obtain accurate TE performance. By the PDOS, we can find the atoms of two outer layers (Te1 and Te3) contribute identically as the inversion symmetry of the structure in α-Te, whereas the contributions of two outer layers (S/Se1 and Te3) are not the same in Janus monolayers. The outer S/Se1 atoms contribute much less than Te2 atoms in the middle layer in Janus monolayers around CBM, while the contributions of outer Te1/Te3 atoms are nearly the same as the Te2 atoms in α-Te.

Table 2. The calculated band gaps with HSE+SOC, HSE, PBE+SOC, and PBE for each monolayer.

|  | HSE+SOC (eV) | HSE (eV) | PBE+SOC (eV) | PBE (eV) |
| --- | --- | --- | --- | --- |
| α-STe$_2$ | 1.22 | 1.51 | 0.78 | 1.05 |
| α-SeTe$_2$ | 0.96 | 1.21 | 0.58 | 0.79 |

| α-Te | 0.70 | 1.12 | 0.44 | 0.76 |

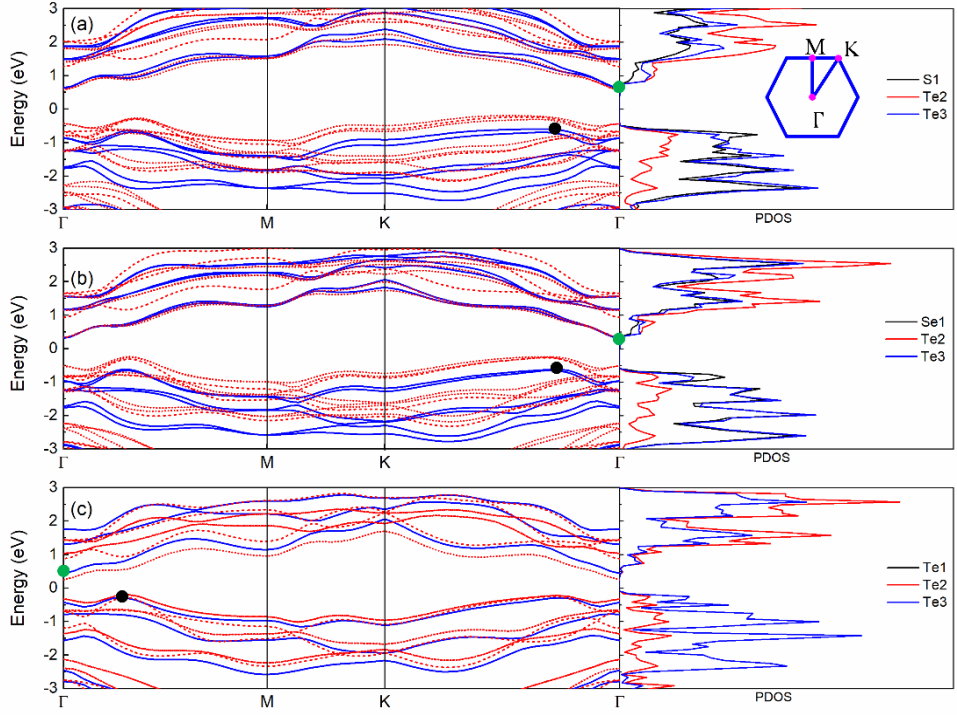

Fig. 2. Electronic band structures of Janus (a) α-STe$_2$, (b) α-SeTe$_2$, and (c) pristine α-Te monolayers with HSE+SOC, exhibited by blue solid lines. For comparison, the band structures HSE without SOC are exhibited by red dashed lines. The partial phonon density of states (PDOS) are also displayed. The inset shows the high-symmetry points chosen. The green and black circles indicate the locations of CBM and VBM of HSE+SOC for each monolayer. The symbols S/Se/Te1, Te2, and Te3 represent atoms of different layers, as shown in Fig. 1.

Table 3. Elastic constant $C_{2D}$, effective mass $m^*$, DP constant $E_i$, carrier mobility $\mu$ and the relaxation time $\tau$ at room temperature. m$_e$ represents the rest mass of the electron.

|  | Carrier type | $C_{2D}$ (N/m) | $m^*$ (m$_e$) | $E_i$ (eV) | $\mu$ (cm$^2$V$^{-1}$s$^{-1}$) | $\tau$ (ps) |
|---|---|---|---|---|---|---|
| α-STe$_2$ | Electron | 50.91 | 0.125 | 6.06 | 1.90×10$^3$ | 0.135 |
| | Hole | | 0.288 | 3.92 | 8.52×10$^2$ | 0.140 |
| α-SeTe$_2$ | Electron | 46.49 | 0.114 | 6.18 | 2.00×10$^3$ | 0.130 |
| | Hole | | 0.288 | 4.36 | 6.30×10$^2$ | 0.103 |
| α-Te | Electron | 40.48 | 0.097 | 5.86 | 2.67×10$^3$ | 0.148 |
| | Hole | | 0.189 | 3.58 | 1.89×10$^3$ | 0.203 |

The phonon dispersions along high-symmetry lines are investigated,

as shown in Fig. 3. The corresponding projected phonon density of states (PhDOS) are also displayed together. There is no negative frequency, indicating the dynamical stability of the three monolayers. In fact, the stabilities of the Janus monolayers are also confirmed in previous works [59,63]. There are three acoustic phonon branches and six optical phonon branches, as there are three atoms in each primitive cell. The out-of-plane acoustic (ZA), transverse acoustic (TA), and longitudinal acoustic (LA) branches are displayed by red, blue and magenta lines, respectively. Note that the ZA phonon branches are quadratic around the Γ point, while TA and LA branches are linear. It is found the highest frequencies of α-STe$_2$, α-SeTe$_2$, and α-Te monolayer are 7.59, 6.23, and 5.73 THz. It mainly results from the fact that α-STe$_2$ has the lightest average atom mass, while the one of α-Te is the heaviest. In α-Te, the Te atom of the middle layer dominates the PhDOS in the high-frequency range of 3.5-6.0 THz, while the atom of the outer layers contributes significantly in the low-frequency range. Moreover, the contributions of Te1 and Te3 are identical as the inversion symmetry of the structure. In Janus monolayers, lighter S/Se atom contributes much more in the high-frequency range. Especially, the S atom is even dominant in the high-frequency range for α-STe$_2$, as it is much lighter than the Te atom. Note there is a phonon gap between acoustic and optical phonon modes in α-Te, whereas it doesn't appear in the two Janus monolayers. Compared to pristine α-Te, the breaking of the symmetry in the Janus monolayers leads to the breaking of degeneracies and obvious splits between phonon branches, such as at the high-symmetry point K.

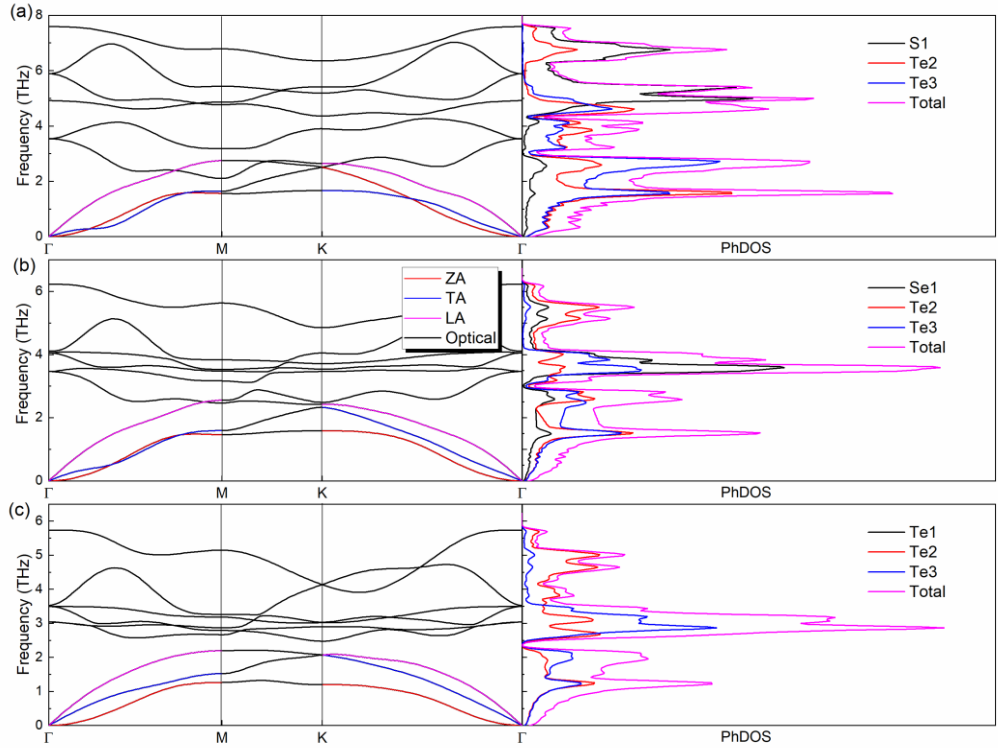

Fig. 3. Phonon dispersions and projected PDOS of (a) α-STe$_2$, (b) α-SeTe$_2$, and (c) α-Te monolayer. Red, blue, and magenta lines indicate ZA, TA, and LA phonon branches, respectively.

The vibrating patterns of the nine phonon modes around Γ point are investigated, as shown in Fig. 4. Note the mode number is given by ascending order in phonon frequency. Mode 1 to 3 belongs to acoustic phonons, while mode 4 to 9 belongs to optical phonons. It is found that modes 4, 5, and 8 of Janus monolayers have significant differences between pristine α-Te monolayer. In α-Te, the vibrations of these modes are inversion symmetry along the c-direction, as the structure is symmetrical. However, as in Janus monolayers, the Te atoms of one outer layer are substituted by S/Se atoms, and the inversion symmetry of the structure is broken, hence the symmetry of vibrating.

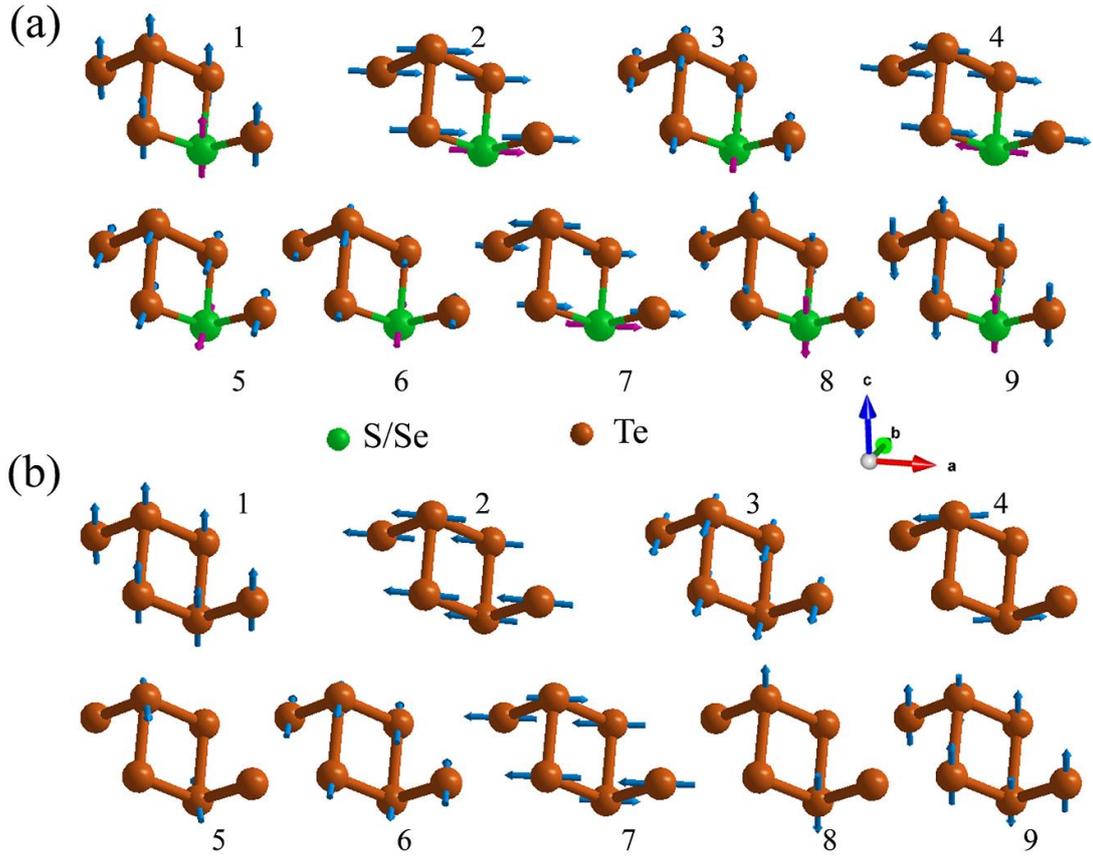

Fig. 4. The vibrating patterns of the nine phonon modes round Γ point for Janus monolayers (a) and pristine α-Te monolayer (b). The arrows show the vibrating directions of each atom. The number is given in ascending order in phonon frequency for each mode.

By solving the Boltzmann transport equation with the constant relaxation time and single parabolic band (SPB) model [64], we investigate the electronic transport coefficients of three monolayers, including the $S$, $\sigma$, as well as PF. Two typical temperatures (300 and 500 K) are selected in the calculation, as shown in Fig. 5. On the whole, it is found the absolute values of Seebeck coefficient $|S|$ decrease for all the three monolayers with the increase of the carrier concentration $n$, as displayed in Fig. 5(a)-(c). Based on the Mahan−Sofo theory [65], $|S|$ for 2D materials can be expressed by the following simple model [12]:

$$|S| = \frac{2\pi^3 k_B^2}{3eh^2 n} m^* T, \qquad (1)$$

where $h$, $k_B$, $m^*$, and $e$ are Planck constant, Boltzmann constant, effective mass, and electron charge, respectively. It is found $|S|$ is proportional to $m^*$, and inversely proportional to $n$ at a given temperature. From Table 3, it is found $m^*$ of the hole is much larger than that of the electron, due to the smaller band curvature around the VBM than the CBM as shown in Fig. 2. Thus, $|S|$ of p-type doping are bigger than the ones of n-type doping for three monolayers. $|S|$ are close to each other among the three monolayers. For instance, at 500 K and $10^{13}$ cm$^{-2}$ of $n$, $|S|$ are 323, 331, and 312 μV/K for α-STe$_2$, α-SeTe$_2$, and α-Te monolayers, respectively. Furthermore, $|S|$ at 500 K is the higher than 300 K, as it is proportional to $T$ based on Eq. (1).

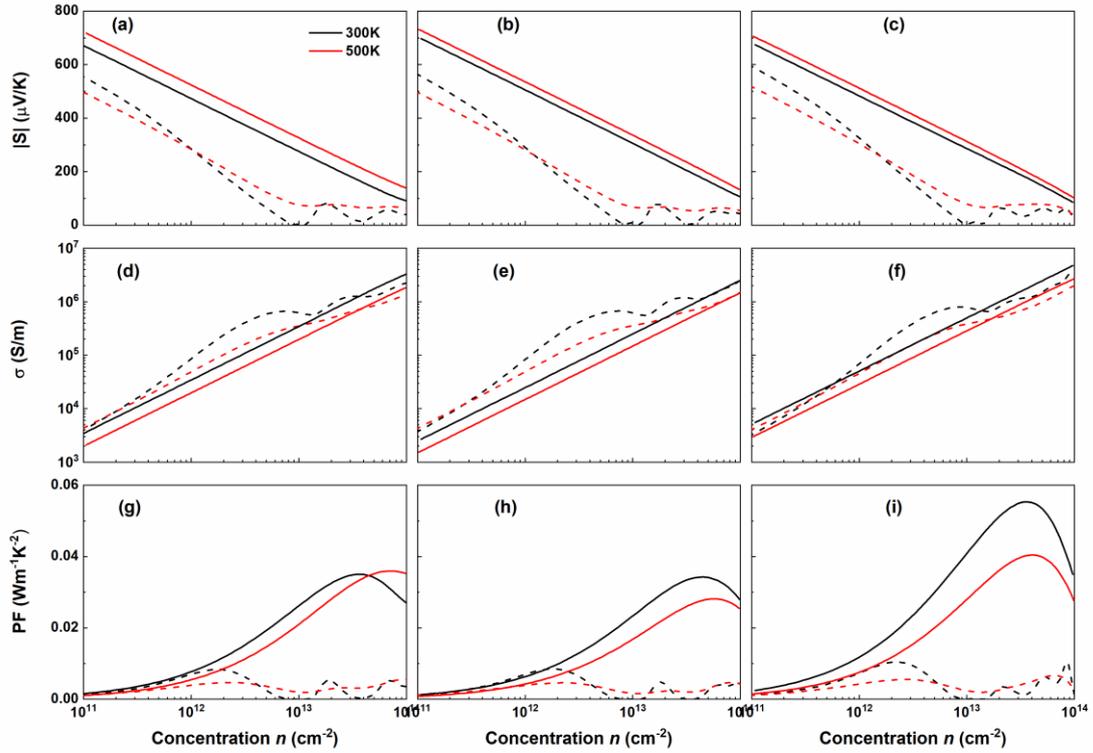

Fig. 5. The concentration $n$ dependences of $|S|$, $\sigma$, and PF at 300 and 500 K for three

monolayers. The solid lines indicate the data of p-type doping, while the dashed lines represent n-type doping. And the 1st, 2nd and 3rd columns are corresponding to the Janus α-STe$_2$, α-SeTe$_2$, and pristine α-Te monolayers, respectively.

The carrier relaxation time $\tau$ is required to obtain $\sigma$ [55]. From the conventional Boltzmann transport theory, electrical conductivity $\sigma$ is proportional to carrier relaxation time $\tau$, which can be expressed as $\tau = \frac{m^*\mu}{e}$, where $\mu$ is the carrier mobility. By using the deformation potential (DP) theory, $\mu$ of 2D materials can be written as [66]:

$$\mu = \frac{e\hbar^3 C_{2D}}{k_B T m^* m_d E_i^2},  \qquad (2)$$

where $\hbar$ and $C_{2D}$ are the reduced Planck constant and 2D elastic constant, respectively. $m_d$ is the average effective mass calculated by $m_d = \sqrt{m_x^* m_y^*}$. However, for in-plane isotropic materials here, $m_x^* = m_y^*$. $E_i = \frac{\Delta E_{edge}}{\Delta a/a}$ is the deformation potential constant, where $\Delta E_{edge}$ is the shift of band energy of VBM / CBM under small compression or expansion, and $\Delta a$ is the change of the lattice parameter relative to the equilibrium lattice parameter $a$. The vacuum energy level correction is considered in the calculations. The calculated data are listed in Table 3. From α-STe$_2$ to α-Te monolayer, $C_{2D}$ becomes smaller, as the S-Te bond is the strongest and the Te-Te bond is the weakest. The carrier relaxation time $\tau$ of electron and hole for three monolayers are close to each other, in the range of 0.1~0.2 ps.

The electrical conductivity $\sigma$ of three monolayers is obtained through the calculated carrier relaxation time $\tau$, as shown in Fig. 4(d)-(f). It is found that $\sigma$ increases with the increasing carrier concentration $n$, and decreases with temperature *T*. The curves of 300 K are higher than the ones of 500 K. On the whole, $\sigma$ of n-type doping is higher than p-type doping at the same *n* and *T*. The dependencies of $\sigma$ can be interpreted by a simple model $\sigma = ne\mu$ [67]. It correlates positively with *n*, contrary to the situation of *S*. Furthermore, $\sigma$ is inversely proportional to $m^*$ and *T* based on Eq. (2). As $m^*$ of the electron is smaller than that of the hole for each monolayer, $\sigma$ of n-type doping is higher.

Then, based on the *S* and $\sigma$, the power factor PF as a function of carrier concentration *n* are calculated and shown in Fig. 4(g)~(i). A high value of PF needs a large *S* and $\sigma$ simultaneously, as $PF = S^2\sigma$. PF has a maximum value at an optimum *n*, as the decreasing function of *S* and the increasing function of $\sigma$. In all cases, the PF value of n-type and p-type doping increases first and then decreases with increasing *n*. It is found PF values of 300 K are generally higher than the values of 500 K for n-type and p-type doping. Moreover, the PF values of n-type doping are much lower than p-type for three monolayers. The maximum values of PF are 0.035, 0.034, and 0.055 W/mK$^2$ for p-type doped α-STe$_2$, α-SeTe$_2$, and α-Te monolayers at 300 K. At 500 K, the maximum values are 0.036, 0.028, and 0.040 W mK$^{-2}$, respectively. It implies the high TE performance of these monolayers, as the high values of PF which can be comparable with many TE materials, such as 2D SnSe and Penta-silicene [12, 68].

A high value of *ZT* also needs a minimum thermal conductivity, including electronic thermal conductivity $\kappa_e$ and lattice thermal conductivity $\kappa_L$. $\kappa_e$ can be calculated through the Wiedemann-Franz law:

$\kappa_e = L\sigma T$, where $L$ is Lorenz number [69]. The calculated $\kappa_e$ of three monolayers at 300 and 500 K are displayed in Fig. 6. Since $\kappa_e$ and $\sigma$ are linearly related, $\kappa_e$ show similar dependences on $n$ and $m^*$. $\kappa_e$ are very small when $n$ is lower than $10^{13}$ cm$^{-2}$, whereas they increase sharply and exceed 15 Wm$^{-1}$ K$^{-1}$ at $10^{14}$ cm$^{-2}$. On the whole, $\kappa_e$ are much larger for n-type doping than p-type doping when $n$ is lower than $10^{13}$ cm$^{-2}$, as $m^*$ of the electron is lighter than the hole. Combined with lower PF, the $ZT$ value of n-type doping should be much smaller than p-type doping. As the $m^*$ of the hole in α-Te monolayer is the lightest among three monolayers, $\kappa_e$ of the α-Te monolayer is the largest. For instance, $\kappa_e$ are 1.30, 0.89, 1.82 Wm$^{-1}$ K$^{-1}$ for n-type doped α-STe$_2$, α-SeTe$_2$, and α-Te monolayers at 300 K, $5 \times 10^{12}$ cm$^{-2}$. It is also found $\kappa_e$ are independent on $T$ approximately.

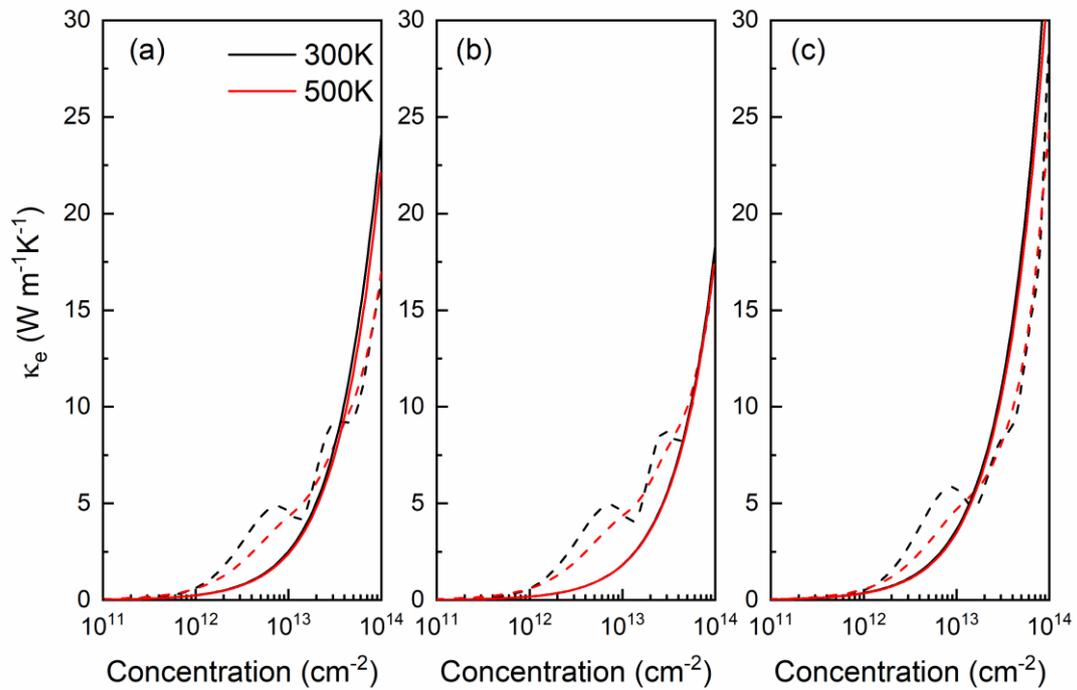

Fig. 6. Electronic thermal conductivity $\kappa_e$ of (a) α-STe$_2$, (b) α-SeTe$_2$, and (c) α-Te monolayers at 300 and 500 K. Solid lines indicate $\kappa_e$ of p-type doping, while dashed lines mean the ones of n-

type doping.

The $\kappa_L$ of full solution for α-STe$_2$, α-SeTe$_2$, and α-Te monolayers are investigated, as shown in Fig. 7. For comparison, the values obtained by the relaxation time approximation (RTA) approach are also displayed. To calculate the $\kappa_L$ of 2D materials, an effective thickness is required, which can be defined as the summation of the buckling height $h$ and the van der Waals radii of the atoms of the outer layer [70]. The effective thicknesses are 7.20, 7.45, and 7.79 Å for α-STe$_2$, α-SeTe$_2$, and α-Te monolayers. The ultralow values of $\kappa_L$ obtained by the full solution (RTA) are 1.18 (0.39), 0.73 (0.32), and 3.01 (1.12) Wm$^{-1}$ K$^{-1}$ at 300 K, respectively. Though the RTA approach underestimates $\kappa_L$ significantly, it is particularly useful for the quite big sample size and slow heating experimental conditions [71-73]. However, our discussion is based on the results of full solution, as our work focuses more on the theoretical research. The values are in agreement with previous works [33, 59]. The $\kappa_L$ are much lower than many 2D monolayers, such as graphene (3716 Wm$^{-1}$ K$^{-1}$) [74, 75], silicene (28.3 Wm$^{-1}$ K$^{-1}$) [74, 75], blue phosphorene (106.6 Wm$^{-1}$ K$^{-1}$) [74], and MoS$_2$ monolayer (87.6 Wm$^{-1}$ K$^{-1}$) [76]. The ultralow $\kappa_L$ generally trend to high TE efficiency. It is also found $\kappa_L$ are close to $\kappa_e$ at 5 × 10$^{12}$ cm$^{-2}$ and room temperature. As well known, $\kappa_L$ usually dominates the TE performance, while the effect of $\kappa_e$ can be neglected. However, $\kappa_e$ are comparable to $\kappa_L$ here, as $\kappa_L$ is quite low (around 1 Wm$^{-1}$ K$^{-1}$). It implies $\kappa_e$ may also play an important role in TE performance. Furthermore, the curves of $\kappa_L$ well satisfy the relationship $\kappa_L \propto 1/T$, indicating the dominant Umklapp process. At 500 K, $\kappa_L$ are reduced to 0.71, 0.45, and 1.83 Wm$^{-1}$ K$^{-1}$ for α-STe$_2$, α-SeTe$_2$, and α-Te monolayers, respectively.

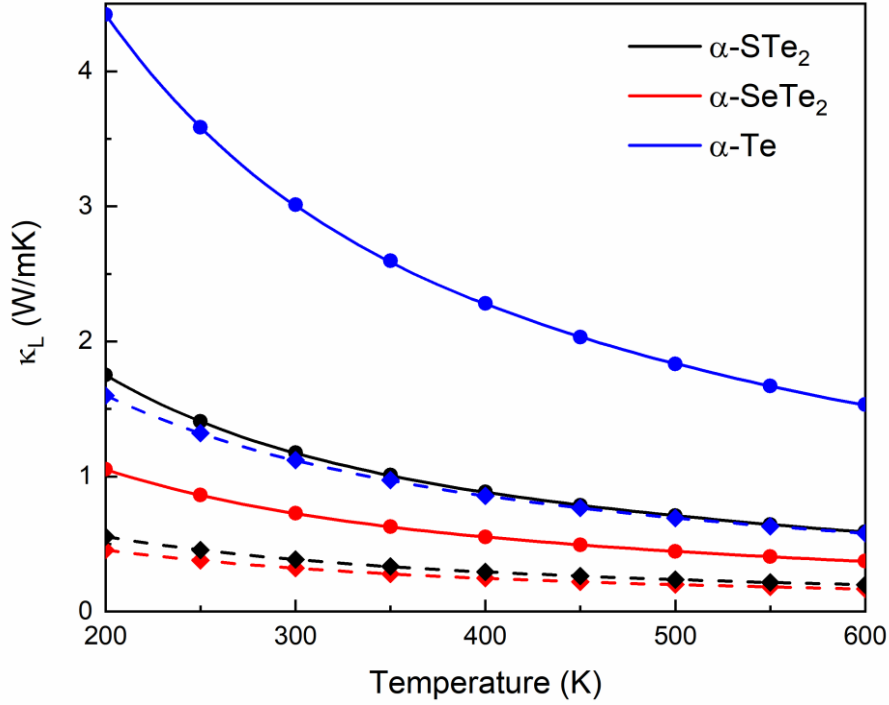

Fig. 7. The temperature dependence of lattice thermal conductivity $\kappa_L$ for α-STe$_2$, α-SeTe$_2$, and α-Te monolayers. The circles indicate for the values of $\kappa_L$ obtained by the full solution, while the solid lines indicate the corresponding 1/T fitting. For comparison, the diamonds and dashed lines are for the data of the RTA approach.

Furthermore, the normalized contribution of each mode to total $\kappa_L$ at 300 K is also investigated, as listed in Table 4. It is found the acoustic modes contribute 67.5%, 42.1% and 57.8% to the total $\kappa_L$ for α-STe$_2$, α-SeTe$_2$, and α-Te monolayers. In other words, the contributions of optical modes are 32.5%, 57.9%, and 42.2%, respectively. The contributions of optical phonons are quite significant, much higher than many other 2D materials, such as graphene [77], MoS$_2$ monolayer [76], and stanene [78].

Table 4. The normalized contribution of each phonon mode to the total $\kappa_L$ for three 2D materials. The mode number is given by ascending order in phonon frequency around Γ point. Mode 1 to 3 belongs to acoustic mode, and the others belong to optical mode.

| Mode | 1 | 2 | 3 | 4 | 5 | 6 | 7 | 8 | 9 |
|---|---|---|---|---|---|---|---|---|---|
| α-STe$_2$ | 0.267 | 0.260 | 0.148 | 0.073 | 0.118 | 0.016 | 0.011 | 0.095 | 0.012 |
| α-SeTe$_2$ | 0.145 | 0.140 | 0.137 | 0.076 | 0.027 | 0.020 | 0.039 | 0.324 | 0.092 |
| α-Te | 0.156 | 0.230 | 0.192 | 0.021 | 0.005 | 0.031 | 0.016 | 0.262 | 0.087 |

Debye temperature $\Theta_D$ is related closely to the thermal properties of materials, which can be expressed as $\Theta_D = h\omega_{max}/k_B$, where $h$ and $\omega_{max}$ is the Planck constant and the maximum of acoustic phonon frequency [79]. They are 131, 123, and 105 K for α-STe$_2$, α-SeTe$_2$, and α-Te monolayers. Usually, high $\Theta_D$ indicates strong harmonic properties, hence high $\kappa_L$. However, α-Te possesses the lowest $\Theta_D$ and the highest $\kappa_L$ among the three monolayers. The abnormal case of $\kappa_L$ stems from $\kappa_L$ is determined by both harmonic and anharmonic properties, whereas $\Theta_D$ measures the harmonic properties only. To unveil the underlying physical mechanisms, the total phase space for three-phonon processes $P_3$ and Grüneisen parameters $\gamma$ are calculated, as displayed in Fig. 8. Note $P_3$ is the direct measure of the number of scattering processes available to each phonon, depending on the phonon dispersions only [75, 80]. Based on Fig. 7(a)-(c), the dispersions of $P_3$ are similar to each other, indicating they are not the main reason for the abnormal $\kappa_L$ of the monolayers. On the other hand, the Grüneisen parameter $\gamma$ is the measure of the anharmonic interactions, and a high value of $\gamma$ leads to a low $\kappa_L$ [75, 81]. Overall, $|\gamma|$ of α-Te is smaller than α-STe$_2$ and α-SeTe$_2$, as shown in Fig. 7(d)-(f), especially for the TA phonon branch. Specifically, the maximum value of $|\gamma|$ is only 2.6 for α-Te, while the ones can reach 20 for α-STe$_2$ and α-SeTe$_2$. The phonon relaxation times $\tau_{ph}$ of three monolayers are also exhibited in Fig. 7(g)-(i). $\tau_{ph}$ of α-STe$_2$ and α-SeTe$_2$ are significantly lower than that of α-Te, resulting from the much higher $|\gamma|$. It is concluded as the breaking of the inversion symmetry with respect to the central plane of the structure, the Janus α-STe$_2$ and α-SeTe$_2$ introduce much stronger anharmonicity, leading to abnormally lower $\kappa_L$ than α-Te [45].

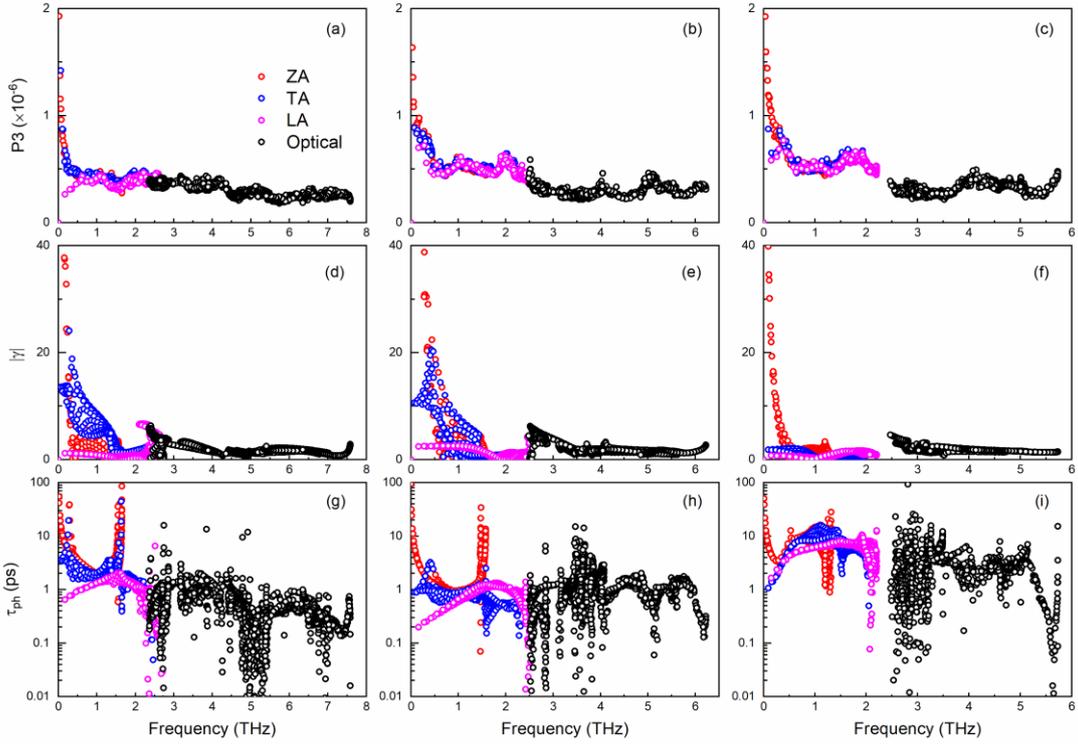

Fig. 8. (a-c) the total phase space for three-phonon processes $P_3$, (d-f) the absolute value of Grüneisen parameters $|\gamma|$, and (g-i) phonon relaxation time $\tau_{ph}$ for the three monolayers. Note the 1st, 2nd and 3rd columns are corresponding to the Janus α-STe$_2$, α-SeTe$_2$ and α-Te monolayers, respectively.

Based on the electronic and thermal transport coefficients, the *ZT* values of three monolayers are estimated and exhibited in Fig. 9. They possess ultrahigh *ZT* values for p-type doping. At 500 K, the three monolayers have the maximum *ZT* values of 3.9, 4.4, and 2.8 for α-STe$_2$, α-SeTe$_2$ and α-Te monolayers, corresponding values of *n* of $4 \times 10^{12}$,

$4 \times 10^{12}$, and $5 \times 10^{12}$ cm$^{-2}$. The α-SeTe$_2$ monolayer has the highest ZT value as the lowest $\kappa_e$ and $\kappa_L$ (0.67 and 0.45 Wm$^{-1}$ K$^{-1}$) at 500 K, while the ones of pristine α-Te are the highest (1.95 and 1.83 Wm$^{-1}$ K$^{-1}$). Therefore, the much lower $\kappa_L$ of α-SeTe$_2$ monolayer mainly leads to a much higher ZT value than α-Te. For the three monolayers studied here, higher PF doesn't always lead to higher ZT. For example, α-Te has the highest PF (0.040 Wm$^{-1}$ K$^{-2}$ at 500 K) and the lowest ZT, whereas α-SeTe$_2$ possesses the lowest PF (0.028 Wm$^{-1}$ K$^{-2}$ at 500 K) and the highest ZT. Thus, it can be concluded Janus structures can enhance TE performance remarkably as they have much lower thermal conductivity, including $\kappa_e$ and $\kappa_L$. Additionally, it is also found $\kappa_e$ are a little higher than $\kappa_L$ in the three monolayers, and can affect ZT significantly. It confirms that $\kappa_e$ can't be neglected for the TE materials with quite low $\kappa_L$. Even at room temperature, these monolayers also have large ZT values, which reach 2.4, 3.1, and 1.9 for α-STe$_2$, α-SeTe$_2$ and α-Te. Thus, these materials are especially suitable for TE devices in daily life. On the other hand, the ZT values are much lower for n-type doping than for p-type, as the smaller PF values and larger $\kappa_e$ values of n-type doping.

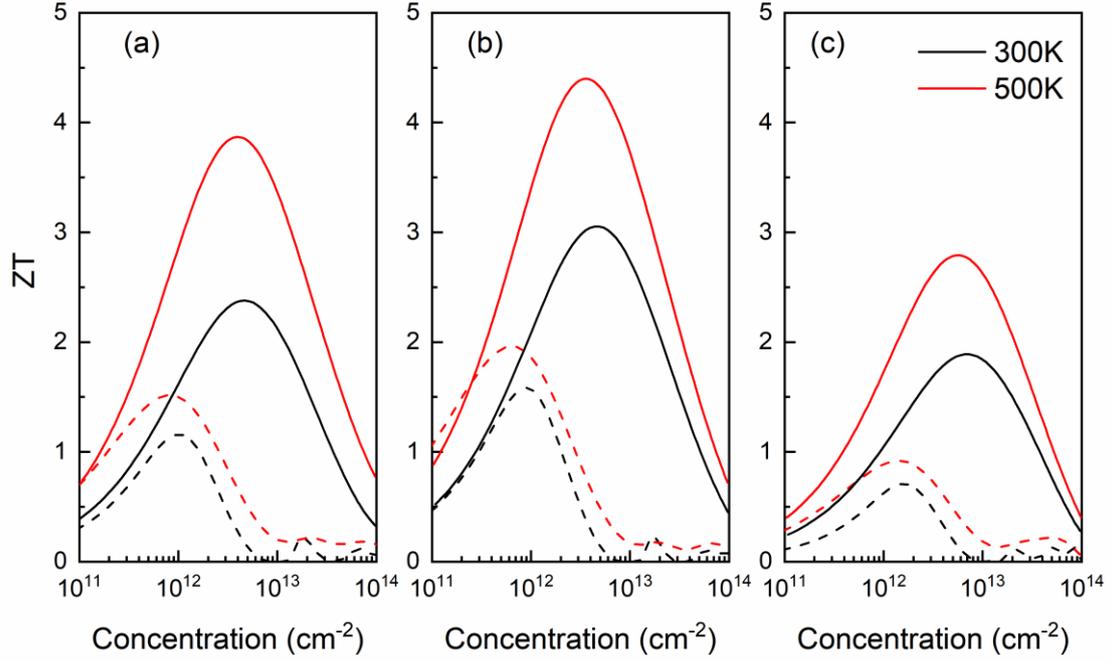

Fig. 9. The figure of merit $ZT$ for (a) Janus α-STe$_2$, (b) α-SeTe$_2$ and (c) pristine α-Te monolayers, respectively. The solid lines indicate the values of p-type doped monolayers, while the dashed lines represent the n-type doping.

## Conclusions

In summary, the electronic, thermal transport properties and TE performances of Janus α-STe$_2$ and α-SeTe$_2$ monolayers are investigated using first-principles calculations combined with semi-classical Boltzmann theory. For comparison, the TE properties of pristine α-Te monolayer are also studied. The electronic band structures, effective masses, and carrier relaxation times are similar to each other. Janus structures have different VBM, larger band gaps, and reduced PF values, compared with the pristine α-Te. However, the breaking of the inversion symmetry greatly reduces the $\kappa_L$ by enhancing the phonon anharmonicity in Janus α-STe$_2$ and α-SeTe$_2$ monolayers. Furthermore, Janus monolayers also possess smaller $\kappa_e$. Note here $\kappa_e$ plays an important role in *ZT* as it can be comparable with $\kappa_L$. Thus, the *ZT* values of Janus α-STe$_2$ and α-SeTe$_2$ monolayers are much higher than the one of the pristine α-Te monolayer. Particularly, the *ZT* value of the Janus α-SeTe$_2$ monolayer is as high as about 4.4 at 500 K. Our work indicates the Janus α-SeTe$_2$ and α-STe$_2$ are promising thermoelectric materials. Furthermore, it also suggests that the Janus structure is an effective method to reduce $\kappa_L$ and enhance *ZT*.

## Acknowledgments

This work was supported by the National Natural Science Foundation of China (Nos. 11974100, 12104356, 61874160) and the Program for Innovative Research Team (in Science and Technology) in University of Henan Province (22IRTSTHN012). Z. Gao acknowledges the support of the China Postdoctoral Science Foundation (No. 2022M712552), and the Fundamental Research Funds for the Central Universities. We also acknowledge the support by HPC Platform, Xi'an Jiaotong University.

## Reference

1    W. Liu, X. Yan, G. Chen and Z. Ren, *Nano Energy*, 2012, **1**, 42-56.
2    J. Yang, L. Xi, W. Qiu, L. Wu, X. Shi, L. Chen, J. Yang, W. Zhang, C. Uher and D. J. Singh, *npj Computational Materials*, 2016, **2**,

15015.

3   T. Zhu, Y. Liu, C. Fu, J. P. Heremans, J. G. Snyder and X. Zhao, *Advanced Materials*, 2017, **29**, 1605884.

4   K. Biswas, J. He, I. D. Blum, C.-I. Wu, T. P. Hogan, D. N. Seidman, V. P. Dravid and M. G. Kanatzidis, *Nature*, 2012, **489**, 414-418.

5   C. M. Jaworski, B. Wiendlocha, V. Jovovic and J. P. Heremans, *Energy & Environmental Science*, 2011, **4**, 4155-4162.

6   X. Su, N. Zhao, S. Hao, C. C. Stoumpos, M. Liu, H. Chen, H. Xie, Q. Zhang, C. Wolverton, X. Tang and M. G. Kanatzidis, *Advanced Functional Materials*, 2019, **29**, 1806534.

7   C. Xiao, K. Li, J. Zhang, W. Tong, Y. Liu, Z. Li, P. Huang, B. Pan, H. Su and Y. Xie, *Materials Horizons*, 2014, **1**, 81-86.

8   N. Wang, M. Li, H. Xiao, Z. Gao, Z. Liu, X. Zu, S. Li and L. Qiao, *npj Computational Materials*, 2021, **7**, 18.

9   M. S. Dresselhaus, G. Chen, M. Y. Tang, R. G. Yang, H. Lee, D. Z. Wang, Z. F. Ren, J. P. Fleurial and P. Gogna, *Advanced Materials*, 2007, **19**, 1043-1053.

10  R. Venkatasubramanian, E. Siivola, T. Colpitts and B. O'Quinn, *Nature*, 2001, **413**, 597-602.

11  Z. Gao, G. Liu and J. Ren, *ACS Applied Materials & Interfaces*, 2018, **10**, 40702-40709.

12  Z. Gao and J. S. Wang, *ACS Applied Materials & Interfaces*, 2020, **12**, 14298-14307.

13  K. S. Novoselov, A. K. Geim, S. V. Morozov, D. Jiang, Y. Zhang, S.

V. Dubonos, I. V. Grigorieva and A. A. Firsov, *Science*, 2004, **306**, 666-669.

14   S. Stankovich, D. A. Dikin, R. D. Piner, K. A. Kohlhaas, A. Kleinhammes, Y. Jia, Y. Wu, S. T. Nguyen and R. S. Ruoff, *Carbon*, 2007, **45**, 1558-1565.

15   A. H. C. Neto and K. Novoselov, *Rep. Prog. Phys.*, 2011, **74**, 082501.

16   K. F. Mak, C. Lee, J. Hone, J. Shan and T. F. Heinz, *Phys. Rev. Lett.*, 2010, **105**, 136805.

17   A. Splendiani, L. Sun, Y. Zhang, T. Li, J. Kim, C.-Y. Chim, G. Galli and F. Wang, *Nano Letters*, 2010, **10**, 1271-1275.

18   L. Li, Y. Yu, G. J. Ye, Q. Ge, X. Ou, H. Wu, D. Feng, X. H. Chen and Y. Zhang, *Nature Nanotechnology*, 2014, **9**, 372-377.

19   C. Zhi, Y. Bando, C. Tang, H. Kuwahara and D. Golberg, *Advanced Materials*, 2009, **21**, 2889-2893.

20   S. Z. Butler, S. M. Hollen, L. Cao, Y. Cui, J. A. Gupta, H. R. Gutiérrez, T. F. Heinz, S. S. Hong, J. Huang, A. F. Ismach, E. Johnston-Halperin, M. Kuno, V. V. Plashnitsa, R. D. Robinson, R. S. Ruoff, S. Salahuddin, J. Shan, L. Shi, M. G. Spencer, M. Terrones, W. Windl and J. E. Goldberger, *ACS Nano*, 2013, **7**, 2898-2926.

21   M. Xu, T. Liang, M. Shi and H. Chen, *Chem. Rev.*, 2013, **113**, 3766-3798.

22   Z. Zhu, X. Cai, S. Yi, J. Chen, Y. Dai, C. Niu, Z. Guo, M. Xie, F. Liu, J. H. Cho, Y. Jia and Z. Zhang, *Phyical Review Letters*, 2017, **119**, 106101.

23	J. Chen, Y. Dai, Y. Ma, X. Dai, W. Ho and M. Xie, *Nanoscale*, 2017, **9**, 15945-15948.

24	Y. Wang, G. Qiu, R. Wang, S. Huang, Q. Wang, Y. Liu, Y. Du, W. A. Goddard, M. J. Kim, X. Xu, P. D. Ye and W. Wu, *Nature Electronics*, 2018, **1**, 228-236.

25	X. Huang, J. Guan, Z. Lin, B. Liu, S. Xing, W. Wang and J. Guo, *Nano Letters*, 2017, **17**, 4619-4623.

26	J. Qin, G. Qiu, J. Jian, H. Zhou, L. Yang, A. Charnas, D. Y. Zemlyanov, C.-Y. Xu, X. Xu, W. Wu, H. Wang and P. D. Ye, *ACS Nano*, 2017, **11**, 10222-10229.

27	A. Apte, E. Bianco, A. Krishnamoorthy, S. Yazdi, R. Rao, N. Glavin, H. Kumazoe, V. Varshney, A. Roy, F. Shimojo, E. Ringe, R. K. Kalia, A. Nakano, C. S. Tiwary, P. Vashishta, V. Kochat and P. M. Ajayan, *2D Materials*, 2018, **6**, 015013.

28	S. Yang, B. Chen, Y. Qin, Y. Zhou, L. Liu, M. Durso, H. Zhuang, Y. Shen and S. Tongay, *Physical Review Materials*, 2018, **2**, 104002.

29	D. Wang, A. Yang, T. Lan, C. Fan, J. Pan, Z. Liu, J. Chu, H. Yuan, X. Wang, M. Rong and N. Koratkar, *Journal of Materials Chemistry A*, 2019, **7**, 26326-26333.

30	Y. Liu, W. Wu and W. A. Goddard, *J. Am. Chem. Soc.*, 2018, **140**, 550-553.

31	J. Qiao, Y. Pan, F. Yang, C. Wang, Y. Chai and W. Ji, *Science Bulletin*, 2018, **63**, 159-168.

32	S. Sharma, N. Singh and U. Schwingenschlögl, *ACS Applied Energy*

*Materials*, 2018, **1**, 1950-1954.

33  J. Ma, F. Meng, J. He, Y. Jia and W. Li, *ACS Appl Mater Interfaces*, 2020, **12**, 43901-43910.

34  Y. C. Cheng, Z. Y. Zhu, M. Tahir and U. Schwingenschlögl, *EPL (Europhysics Letters)*, 2013, **102**, 57001.

35  A.-Y. Lu, H. Zhu, J. Xiao, C.-P. Chuu, Y. Han, M.-H. Chiu, C.-C. Cheng, C.-W. Yang, K.-H. Wei, Y. Yang, Y. Wang, D. Sokaras, D. Nordlund, P. Yang, D. A. Muller, M.-Y. Chou, X. Zhang and L.-J. Li, *Nature Nanotechnology*, 2017, **12**, 744-749.

36  J. Zhang, S. Jia, I. Kholmanov, L. Dong, D. Er, W. Chen, H. Guo, Z. Jin, V. B. Shenoy, L. Shi and J. Lou, *ACS Nano*, 2017, **11**, 8192-8198.

37  R. Li, Y. Cheng and W. Huang, *Small*, 2018, **14**, 1802091.

38  X. Yang, D. Singh, Z. Xu, Z. Wang and R. Ahuja, *Journal of Materials Chemistry C*, 2019, **7**, 12312-12320.

39  X. Yang, A. Banerjee and R. Ahuja, *ChemCatChem*, 2020, **12**, 6013-6023.

40  A. Rawat, M. K. Mohanta, N. Jena, Dimple, R. Ahammed and A. De Sarkar, *The Journal of Physical Chemistry C*, 2020, **124**, 10385-10397.

41  X. Yang, D. Singh, Z. Xu and R. Ahuja, *New Journal of Chemistry*, 2020, **44**, 7932-7940.

42  S.-D. Guo, *Physical Chemistry Chemical Physics*, 2018, **20**, 7236-7242.

43  S.-D. Guo, X.-S. Guo and Y. Deng, *Journal of Applied Physics*, 2019,

**126**, 154301.

44  L. Cao, Y. S. Ang, Q. Wu and L. K. Ang, *Applied Physics Letters*, 2019, **115**, 241601.

45  R. Gupta, B. Dongre, C. Bera and J. Carrete, *The Journal of Physical Chemistry C*, 2020, **124**, 17476-17484.

46  A. Patel, D. Singh, Y. Sonvane, P. B. Thakor and R. Ahuja, *ACS Appl Mater Interfaces*, 2020, **12**, 46212-46219.

47  G. Kresse and D. Joubert, *Physical Review B*, 1999, **59**, 1758-1775.

48  G. Kresse and J. Furthmüller, *Physical Review B*, 1996, **54**, 11169-11186.

49  J. P. Perdew, K. Burke and M. Ernzerhof, *Phys. Rev. Lett.*, 1996, **77**, 3865-3868.

50  H. J. Monkhorst and J. D. Pack, *Physical Review B*, 1976, **13**, 5188-5192.

51  S. Grimme, *Journal of Computational Chemistry*, 2006, **27**, 1787-1799.

52  J. Heyd, G. E. Scuseria and M. Ernzerhof, *The Journal of chemical physics*, 2003, **118**, 8207-8215.

53  W. G. Hoover, *Phys. Rev. A*, 1985, **31**, 1695-1697.

54  S. Nosé, *Molecular Physics*, 1986, **57**, 187-191.

55  G. K. H. Madsen, J. Carrete and M. J. Verstraete, *Computer Physics Communications*, 2018, **231**, 140-145.

56  A. Togo and I. Tanaka, *Scripta Materialia*, 2015, **108**, 1-5.

57  J. Carrete, B. Vermeersch, A. Katre, A. van Roekeghem, T. Wang, G.

K. H. Madsen and N. Mingo, *Computer Physics Communications*, 2017, **220**, 351-362.

58	V. B. Anzin, M. I. Eremets, Y. V. Kosichkin, A. I. Nadezhdinskii and A. M. Shirokov, *physica status solidi (a)*, 1977, **42**, 385-390.

59	J. Singh, M. Jakhar and A. Kumar, *Nanotechnology*, 2022, **33**, 215405.

60	H. Usui, K. Suzuki, K. Kuroki, S. Nakano, K. Kudo and M. Nohara, *Physical Review B*, 2013, **88**, 075140.

61	K. Kuroki and R. Arita, *J. Phys. Soc. Jpn.*, 2007, **76**, 083707.

62	H. Usui and K. Kuroki, *Journal of Applied Physics*, 2017, **121**, 165101.

63	Y. Chen, J. Liu, J. Yu, Y. Guo and Q. Sun, *Phys Chem Chem Phys*, 2019, **21**, 1207-1216.

64	M. Cutler, J. F. Leavy and R. L. Fitzpatrick, *Physical Review*, 1964, **133**, A1143-A1152.

65	G. D. Mahan and J. O. Sofo, *Proceedings of the National Academy of Sciences*, 1996, **93**, 7436-7439.

66	J. Qiao, X. Kong, Z.-X. Hu, F. Yang and W. Ji, *Nature Communications*, 2014, **5**, 4475.

67	G. J. Snyder and E. S. Toberer, *Nature Materials*, 2008, **7**, 105-114.

68	F. Q. Wang, S. Zhang, J. Yu and Q. Wang, *Nanoscale*, 2015, **7**, 15962-15970.

69	G. K. H. Madsen and D. J. Singh, *Computer Physics Communications*, 2006, **175**, 67-71.

70  X. Wu, V. Varshney, J. Lee, Y. Pang, A. K. Roy and T. Luo, *Chem. Phys. Lett.*, 2017, **669**, 233-237.

71  P. Torres, F. X. Alvarez, X. Cartoixà and R. Rurali, *2D Materials*, 2019, **6**, 035002.

72  B. Lv, X. Hu, N. Wang, J. Song, X. Liu and Z. Gao, *Applied Surface Science*, 2021, **559**, 149463.

73  N. Wang, H. Gong, Z. Sun, C. Shen, B. Li, H. Xiao, X. Zu, D. Tang, Z. Yin, X. Wu, H. Zhang and L. Qiao, *ACS Applied Energy Materials*, 2021, **4**, 12163-12176.

74  B. Peng, D. Zhang, H. Zhang, H. Shao, G. Ni, Y. Zhu and H. Zhu, *Nanoscale*, 2017, **9**, 7397-7407.

75  B. Peng, H. Zhang, H. Shao, Y. Xu, G. Ni, R. Zhang and H. Zhu, *Physical Review B*, 2016, **94**, 245420.

76  B. Peng, H. Zhang, H. Shao, Y. Xu, X. Zhang and H. Zhu, *Annalen der Physik*, 2016, **528**, 504-511.

77  L. Lindsay, W. Li, J. Carrete, N. Mingo, D. A. Broido and T. L. Reinecke, *Physical Review B*, 2014, **89**, 155426.

78  B. Peng, H. Zhang, H. Shao, Y. Xu, X. Zhang and H. Zhu, *Scientific Reports*, 2016, **6**, 20225.

79  L.-D. Zhao, S.-H. Lo, Y. Zhang, H. Sun, G. Tan, C. Uher, C. Wolverton, V. P. Dravid and M. G. Kanatzidis, *Nature*, 2014, **508**, 373-377.

80  W. Li and N. Mingo, *Physical Review B*, 2015, **91**, 144304.

81  Z. Gao, F. Tao and J. Ren, *Nanoscale*, 2018, **10**, 12997-13003.